\documentclass[12pt]{article}
\usepackage[latin1]{inputenc}
\usepackage{graphicx}
\newcommand{\el}{\ell}

\newcommand{\text}{\rm}

\newcommand{\ug}{ \; = \; }

\newcommand{\bb}{\begin{equation}}
\newcommand{\ee}{\end{equation}}
\newcommand{\bega}{\begin{eqnarray}}
\newcommand{\ega}{\end{eqnarray}}
\newcommand{\begae}{\begin{eqnarray*}}
\newcommand{\egae}{\end{eqnarray*}}

\newcommand{\h}{\hspace*{4ex}}
\newcommand{\dis}{\displaystyle}

\newcommand{\om}{\omega}

\newcommand{\cent}{\centerline}
\newcommand{\vs}{\vspace*}

\begin{document}

\baselineskip 0.8cm

\begin{center}

{\large {\bf Structured Light by linking together
diffraction-resistant spatially shaped beams: ``Lego-beams''} \ ${\;^\dag }$ }
\footnotetext{$^{\: ^\dag}$  Work partially supported by
FAPESP, CAPES and CNPq. \ E-mail addresses for contacts: mzamboni@decom.fee.unicamp.br; michelunicamp@yahoo.com.br [MZR]}


\end{center}

\vs{5mm}

\cent{ Michel Zamboni-Rached$^{\rm a,b}$, Erasmo Recami$^{\rm
a,c,d} \; (^*)$,  Tarcio A.Vieira$^{\rm e}$, Marcos
R.R.Gesualdi$^{\rm e}$,  and J.N. Pereira$^{\rm a}$}

\footnotetext{$^{\:(^*)}$ Visiting c/o Decom, Unicamp, by a PVE fellowship of CAPES (Brazil)}

\vspace*{0.2 cm}

\cent{{\em $^{\rm a}$DECOM--FEEC, \ UNICAMP, \ Campinas, S.P., Brasil}}

\cent{{\em $^{\rm b}$University of Toronto, Toronto, ON, Canada}}

\centerline{{\em $^{\rm c}$Facolt\`{a} di Ingegneria,
Universit\`{a} statale di Bergamo, Dalmine (BG), Italy}}

\centerline{{\em $^{\rm d}$INFN---Sezione di Milano, Milan,
Italy}}

\centerline{{\em $^{\rm e}$CECS, Federal University of ABC, CP
09210-580, Santo Andre, SP, Brazil}} \vs{5mm}

%
%

\vs{0.5 cm}

{\bf Abstract  \ --} \ In this paper we present a theoretical
method, together with its experimental confirmation, to obtain
structures of light by connecting diffraction-resistant
cylindrical beams of finite lengths and different radii. The
resulting ``Lego-beams'' can assume, on demand, various
unprecendent spatial configurations. We also experimentally
generate some of them on using a computational holographic
technique and a spatial light modulator. Our new, interesting
method of linking together various {\em pieces of light} can find
applications in all fields where structured light beams are
needed, in particular such as optical tweezers, e.g. for
biological manipulations, optical guiding of atoms, light orbital
angular momentum control, holography, lithography,
non-linear-optics, interaction of electromagnetic radiation with
Bose-Einstein condensates, and so on, besides in general the field
of Localized Waves (non-diffracting beams and pulses).



\section{Introduction}

\h Structured Light \cite{fw1,fw2,fw3,livro1,livro2,livro3} has
been more and more studied, and applied in various sectors, like
optical tweezers
\cite{ashkin1,arlt,chavez1,macdonald,chavez2,mcgloin,grier,padg,leo1,dolak1,dolak2},
optical guiding of atoms
\cite{wang,lembe,arlt2,rhodes,court,dolak3,edwin}, imaging
\cite{imaging}, light orbital angular momentum control and
applications \cite{ahmed1,allen1,andrews1,araoka,chavez3,livro4},
and photonics in general.

\h A rather efficient method to model longitudinally the intensity
of non-diffracting beams is by the so-called Frozen Waves (FWs)
\cite{fw1,fw2,fw3,tarcio1,tarcio2,mateus,fw4,ahmed2}, obtained
from superpositions of co-propagating Bessel beams, endowed with
the same frequency and order. The resulting diffraction resistant beam, with a
longitudinal intensity shape freely chosen {\em a priori}, may
then propagate, remaining confined, along the propagation axis
$z$, or over a cylindrical surface (depending on the order of the
constituting Bessel beams), while its ``spot" size, and the
cylindrical surface radius, can be as well chosen {\em a priori}.
In this way, it is possible to construct, e.g., cylindrical beams
whose static envelopes possess non-negligible energy density in
finite, well-defined spatial intervals only: so that they can be
regarded as {\em segments or cylindrical pieces of light}.

\h Aiming also at a greater control on the beam {\em transverse}
shape, another method was recently proposed \cite{edwin,ahmed1},
where different-order FW-type beams are superposed, which possess
appreciable intensities along different, but consecutive, space
intervals: So that one ends with cylindrical structures of light
endowed with different radii and located in different positions
along the $z$ axis. This new method resulted efficient,
incidentally, also for controlling the orbital angular momentum
along the propagation axis \cite{ahmed1}.

\h Anyway, and interestingly enough, it is possible to join
together in the same way even two FW-type beams {\em bearing the
same order}, by getting  again a structure with two
different-radius cylinders, each one in its own space interval. To
this aim, it is sufficient that each equal-order FW possesses a
different value of its central longitudinal wave number: which
implies a different radius for the corresponding cylindrical
structure. An advantage of using FWs with the same order is that
the resulting beam intensity keeps its azimuthal symmetry, thus
avoiding the intensity perturbations (asymmetries) that one meets
on the transverse plane where two different-order FWs connect to
each other.

\h Indeed, in this work we extend --theoretically and
experimentally-- on a method of ours proposed in
\cite{edwin,ahmed1}, by explicitly considering superpositions not
only of different order FWs, but also of FWs with the same order
but different central longitudinal wavenumbers. Moreover, we also
investigate the superposition of FWs whose longitudinal intensity
values have non-zero values inside {\em the same} space interval:
This allows us creating even coaxial light structures, or
cylindrical structures e.g. with ``emboli'' (blockages), and so on.

\h We call ``Lego-beams" all such structures of light. Our new
method of linking together various ``pieces of light" can find
applications in all fields where structured light beams are
needed, such as in particular optical tweezers, e.g. for
biological manipulations, optical guiding of atoms, light orbital
angular momentum control, holography, lithography,
non-linear-optics, interaction of electromagnetic radiation with
Bose-Einstein condensates, and so on, besides, of course, the
field of Localized Waves (non-diffracting beams and pulses).

\section{The method}

\h Let us consider as exact solutions to the wave equation the
following superposition:

\bb \Psi(\rho,\phi,z,t) \ug \sum_{\nu =
-\infty}^{\infty}\,\sum^{\infty}_{\el=-\infty}\psi_{\nu
\el}(\rho,\phi,z,t) \label{lb} \ee

with

\bb \psi_{\nu \el}(\rho,\phi,z,t) \ug \mathcal{M}_{\nu}\,
\sum^{N_{\nu \el}}_{n=-N_{\nu \el}}A_{\nu \el n}
        J_{\nu}(h_{\nu \el n}\rho)e^{i \nu \phi}e^{\beta_{\nu \el n} z} \label{FW}   \ee

where

\bb  \beta_{\nu \el n} \ug Q_{\nu \el} + \frac{2\pi}{L}n \,\,\, ,
\label{kz} \ee

\bb  h_{\nu \el n} \ug \sqrt{k^2 - \beta_{\nu \el n}^2 } \,\,\, ,
\label{h} \ee

\bb A_{\nu \el n} \ug \frac{1}{L} \int_{0}^{L} F_{\nu
\el}(z)e^{-i\frac{2\pi}{L}n z} dz \,\,\, , \label{An}  \ee

and with \ $k = \om/c$ \ and \ $\mathcal{M}_{\nu} =
1/[J_{\nu}(.)]_{max}$, \ quantity \ $[J_{\nu}(.)]_{max}$ \ being
the maximum value of the $\nu$-order Bessel function of the first
kind.

\h For a better comprehension of it, let us explicitly examine the
solution given by Eq.(\ref{lb}). First, let $\nu$ e $\el$ have
fixed values, and then consider a term $\psi_{\nu
\el}(\rho,\phi,z,t)$ of the series. Eq.(\ref{FW}) teaches us that
such term is a FW \cite{fw1,fw2,fw3} of order $\nu$, whose central
longitudinal wavenumber ($n=0$) is \ $\beta_{\nu \el 0} = Q_{\nu
\el}$, so that $h_{\nu \el 0} \ug \sqrt{k^2 - Q_{\nu \el}^2 }$;
and the FW intensity shape $|F_{\nu \el}(z)|^2$ in the interval $0
\leq z \leq L$ results to be concentrated: (i) either on a
cylindrical surface having radius $\rho_{\nu}\approx s_{\nu}/
h_{\nu \el 0}$ [quantity $s_{\nu}$ being the $s$ value where
$J_{\nu}$(s) gets its maximum value], in the case $\nu \geq 1$; \
or \ (ii) around the propagation axis $z$, with a spot radius
$r_{0}\approx 2.4/ h_{0 \el 0}$, in the case $\nu =0$. Here, we
define $\rho_{0} \equiv 0$. In any case, $|\psi_{\nu
\el}(\rho=\rho_{\nu},\phi,z,t)|^2 \approx |F_{\nu \el}(z)|^2$.

\h Let us now examine the sum $\sum_{\el}\psi_{\nu \el}$ entering
Eq.(\ref{lb}). It represents a superposition of FWs of the same
order ``$\nu$'', but with different values of their central
longitudinal wavenumbers $\beta_{\nu \el 0} \ug Q_{\nu \el}$. Each
FW (with the same order, let us repeat, but different $\el$)
possesses its own intensity longitudinal shape  $|F_{\nu
\el}(z)|^2$.

\h Finally, the last sum, that is, $\sum_{\nu}\sum_{\el}\psi_{\nu
\el}$, refers to superposition of {\em different order} FWs: as
before, for each ``$\nu$'' we deal with FWs with different
$\el$ values, that is, possessing different values of $Q_{\nu
\el}$ for their central longitudinal $\beta_{\nu \el 0}$ and
also possessing their own longitudinal intensity shapes.

\h Before going on, let us stress the type of light structures
which can be properly built up by the solution we started from. We
know that each $\psi_{\nu \el}$ in superposition (\ref{lb}) is a
FW whose \emph{static intensity envelope} can be regarded as a ``piece''
of light.  Due to the interference present in any wave phenomena,
when summing different FWs together, like $\psi_1$ and $\psi_2$,
one does not obtain the mere sum of their intensity envelopes. For
instance, as well-known,

\bb |\psi_1 + \psi_2|^2 = |\psi_1|^2 +
|\psi_2|^2 + 2|\psi_1||\psi_2|\cos\delta \ ,
\label{Dp} \ee

where $\delta$ is the phase difference between the two waves.  To
minimize the contribution associated with these phase differences,
one has to avoid e.g. cases in which two different FWs have
relevant intensities in the same spatial region. We can choose,
for instance, different FWs, $\psi_{\nu \el}$, which in the region
$0 \leq z \leq L$ have appreciable longitudinal intensity
patterns, $|F_{\nu \el}(z)|^2$, only in different, and
consecutive, intervals. Even when minimizing the interference, it
goes on existing; its critical effect being in the welding region:
In other words, critical interference takes mainly place in the
$z$-planes corresponding to the interval boundaries. A way out is
choosing successive FWs with orthogonal polarizations, so that the
double product in the last equation does vanish: This will be
examined in future works.

\h Actually, it is also possible to obtain interesting light structures
by superposing FWs with non-zero intensities over the same
$z$-interval, provided that the corresponding cylindrical
structures have {\em rather different} radii: Thus, obtaining
interesting co-axial structures for light.

\h Everything results clearer from the theoretical, and
experimental, examples presented in the next Section.

\

\section{Theoretical Examples}

In this section we apply our method to obtain some interesting
{\em structured} optical beams. We adopt a frequently used
wavelength: $\lambda = 632.8\;$nm.

\h We are going to represent with enough details our {\em first
example} of a ``Lego-beam'': Namely, of a
beam whose spatial structure consists of two adjacent cylindrical
surfaces, $12$cm and $16$cm long, with corresponding radii of
$128\mu$m and $182\mu$m subsequently linked each other; besides a
coaxial, central light-segment having spot $15\mu$m and length
$28$cm.

\h To get such a beam, we use our ``Lego-beam''-type solution,
Eq.(\ref{lb}), and consider only three FWs, one of order zero, and
two more of the same order 10 (but with different values of their
central longitudinal wave numbers). That is, by inserting in
Eq.(\ref{lb}) the non-zero functions (FWs)  $\psi_{0\,0}$, \
$\psi_{10\,0}$ \ and \ $\psi_{10\,1}$, endowed in the interval $0
\leq z \leq L = 1\;$m with the following longitudinal intensity
shapes:

\bb \begin{array}{clr} F_{\nu l} \,\, =&
\delta_{\nu\,0}\,\delta_{\el\,0}\,[H(z-0.2)-H(z-0.48)] \\
\\
& + 0.8\delta_{\nu\,10}\,\delta_{\el\,0}\,[H(z-0.2)-H(z-0.325)] \\
\\
& + 0.8\delta_{\nu\,10}\,\delta_{\el\,1}\,[H(z-0.32)-H(z-0.48)]
\,\, ,
\end{array} \label{F1} \ee

where $\delta_{p\,q}$ is the Kronecker delta function and $H(.)$
is the Heaviside function. \ The corresponding $Q_{\nu \el}$ are
evaluated on the basis of the wished values for the cylindical
surface radii and for the light-segment spot, resulting to be:

\bb \begin{array}{clr} & Q_{0\,0} = 0.9999100\dis{\frac{\om}{c}} \\
\\
& Q_{10\,0} = 0.9999700\dis{\frac{\om}{c}} \\
\\
& Q_{10\,1} = 0.9999850\dis{\frac{\om}{c}}
\end{array} \label{F1} \ee

\h Then, the coefficients $A_{\nu \el n}$ can be easily obtained
via Eq.(\ref{An}); while the longitudinal and transverse wave
numbers,  $\beta_{\nu \el n}$ \ and \ $h_{\nu \el n}$,
respectively, are got by Eqs(\ref{kz},\ref{h}). In the present
example, it was enough to use 25 Bessel beams in each one of the
considered FWs: that is, $N_{\nu \el} = 12$.

\h Our result is represented by the following Figures.  First,
Figures 1(a), 1(b) and 1(c) show, separately, the intensities of the
three FWs which are going to compose our beam (in a sense, show
the pieces of the ``Lego-beam''). Figure 2 shows the intensity,
instead, of the resulting beam (that is, of the structured
light resulting from the sum of the chosen FWs).

\h One can notice that the interferences among the three FWs got
minimized by the fact that such FWs possess relevant intensities
in different space regions.

\begin{figure}[!h]
\begin{center}
 \scalebox{.17}{\includegraphics{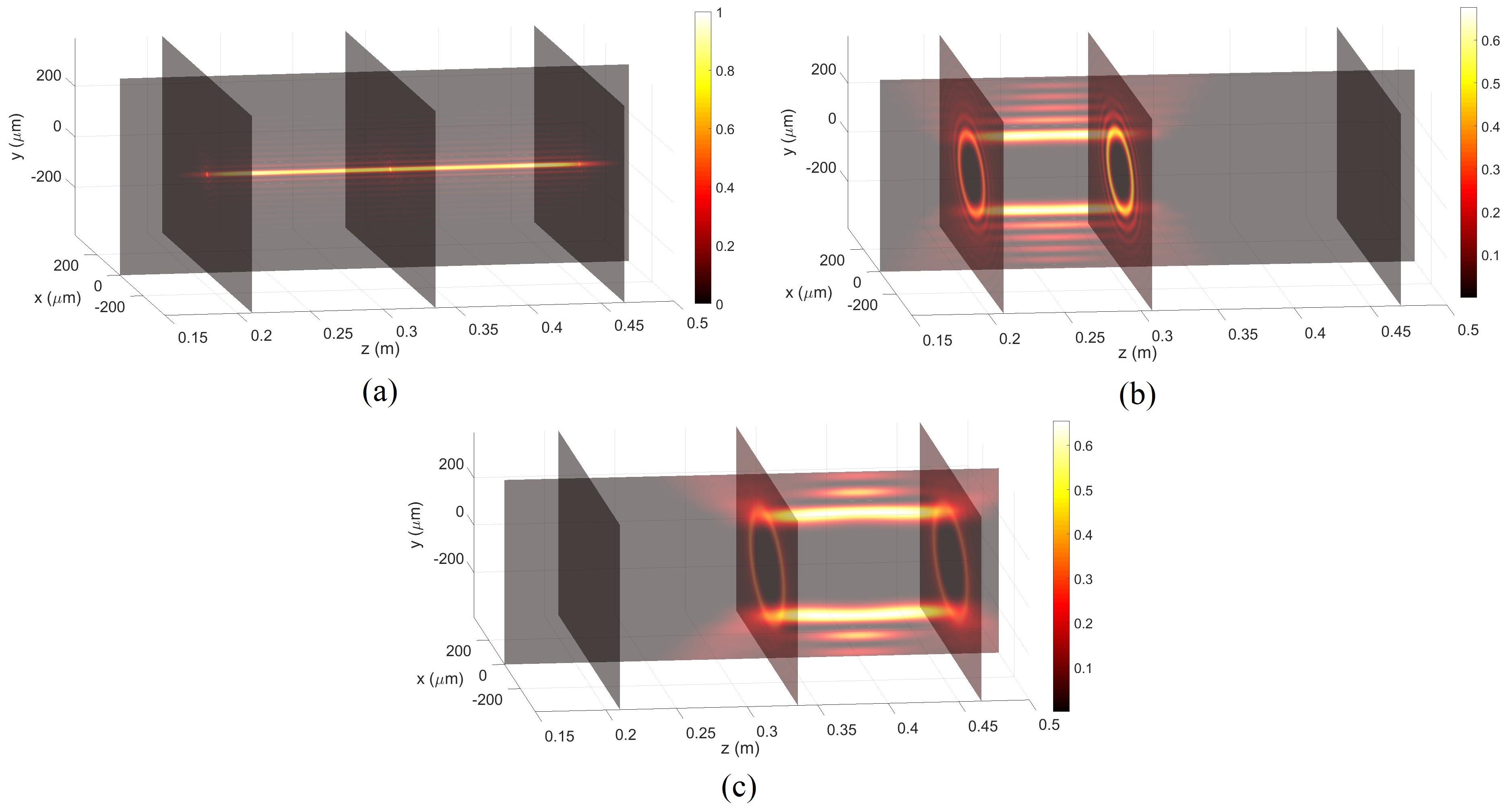}}
\end{center}
\caption{Figures (a), (b) and (c) show, separately, the
intensities of the three FWs which are going to compose our beam
(in a sense, show the pieces of the ``Lego-beam'')}
\label{fig1}
\end{figure}

\begin{figure}[!h]
\begin{center}
 \scalebox{.45}{\includegraphics{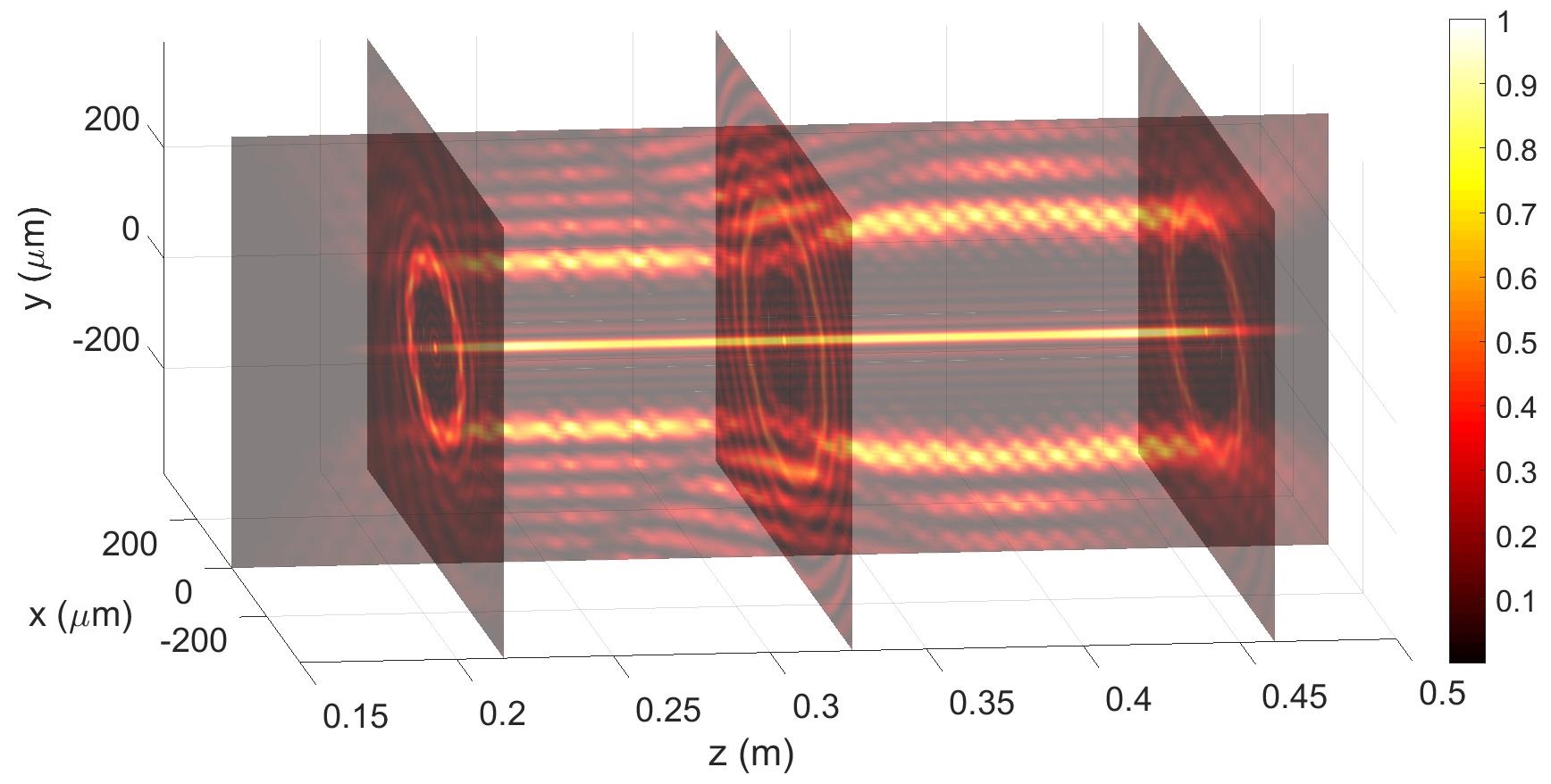}}
\end{center}
\caption{The intensity is here shown of the resulting ``Lego-beam'' (that
is, of the structured light resulting from the sum of the chosen FWs) }
\label{fig2}
\end{figure}

\newpage

\begin{figure}[!h]
\begin{center}
 \scalebox{.27}{\includegraphics{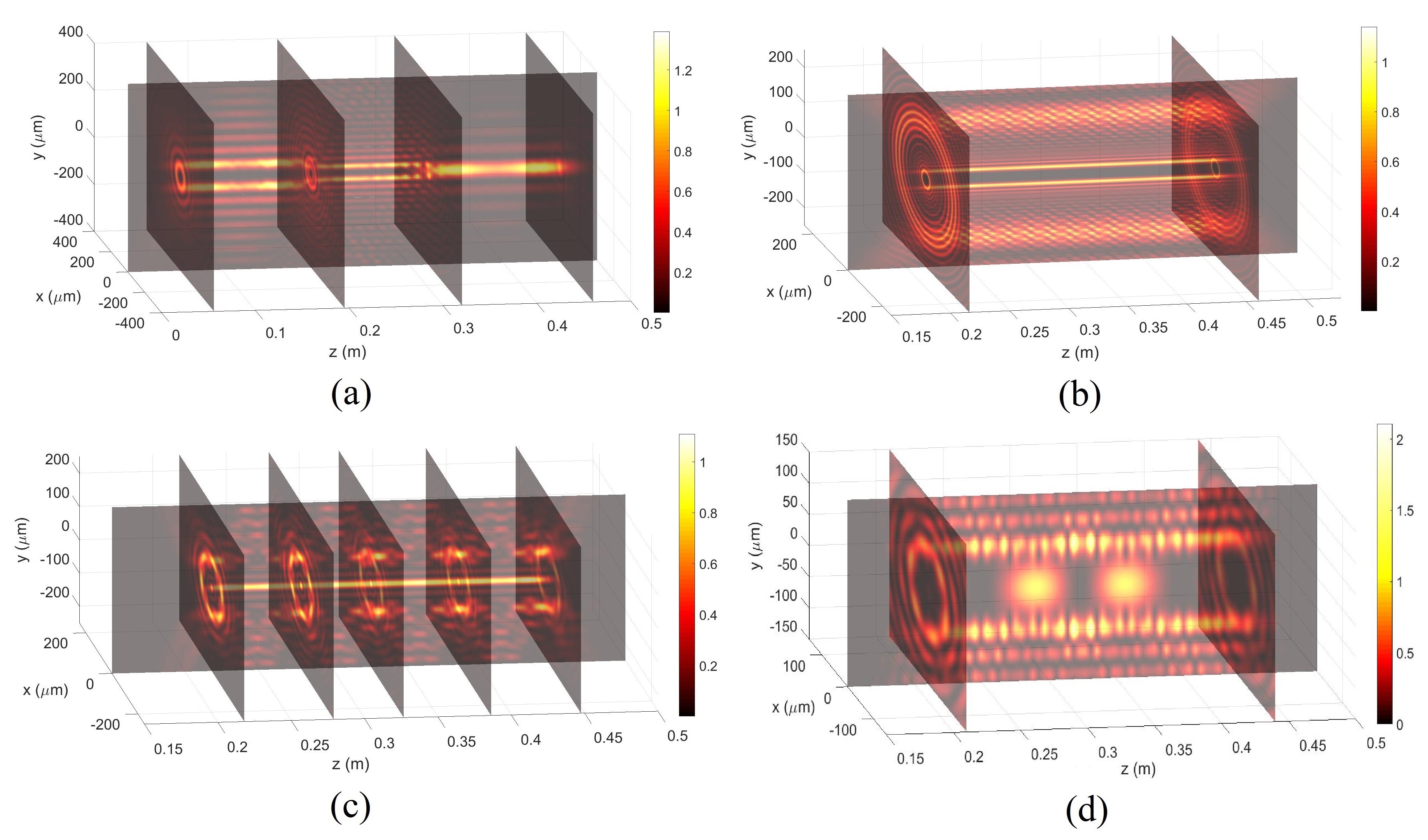}}
\end{center}
\caption{Figures (a), (b), (c) and (d) depict four {\em further}
interesting ``Lego-beams'' obtained by our method. They refer to
cylindrical or threadlike structures (adjacent or coaxial in their
location), having transverse sizes of tens of micrometers and
longitudinal sizes of tens of centimeters: They being, therefore,
highly resistant to diffracting effects. \ See also the text.} \label{fig3}
\end{figure}

\h Let us show some further possibilities forwarded by our method,
even if --for the sake of conciseness-- we skip mathematical
details: Namely, let us show in Figs.3 four further {\em light
structures} obtained from Eq.(\ref{lb}).

\h The first one, Fig.3(a), depicts the intensity of the
``Lego-beam'' consisting in two cylindrical surfaces, of different
(decreasing) radii, linked one to the other, while the second
cylinder on its right side has a light segment acting as a
``cork". \ Figure 3(b) shows the intensity of a structured light
beam formed by two subsequent, coaxial cylindrical surfaces: In
this case two FWs with the same order (larger than zero, of
course) were adopted. \  In Fig.3(c) one has a sequence of
donut-shaped light structures, with a central coaxial (zero order)
light line. \ At last, Fig.3(d) refers to the intensity of a
structured beam, made of a cylindrical surface with two light
``embuli'' (blockages) inside it.

\h All these examples refer to cylindrical or threadlike
structures (adjacent or coaxial in their location), having
transverse sizes of tens of micrometers and longitudinal sizes of
tens of centimeters: They being, therefore, highly resistant to
diffracting effects.

\section{Experimental Confirmations}

\h Let us present in this Section a couple of experimental
confirmations of our approach, by constructing two ``Lego-beams''
via a holographic setup that performs  the optical reconstruction
of computer generated holograms, sent electronically into a
reflective Spatial Light Modulator (SLM), used in amplitude modulation
mode, followed by a 4F spatial filtering system. Further details can
be found in the caption of Fig.4.

\h We obtain the 2D CGH of a ``Lego-beam'' through the complex
transmittance function (hologram) obtained from the desired
resulting beam at the origin $\Psi(\rho,\phi,z=0,t)$, \ with
$\Psi$ given by Eq.(\ref{lb}). The hologram equation is expressed
by \cite{tarcio1,tarcio2,arrizon1,arrizon2} :

\bb H(x,y) \ug \frac{1}{2}\left\{\beta(x,y) +
\alpha(x,y)\cos\left[\theta(x,y) - 2\pi(u_0x+v_0y)\right] \right\}
\ee

where \ $\alpha(x,y)$ \ and \ $\theta(x,y)$ \ are amplitude and
phase, respectively, of the complex field \
$\Psi(\rho,\phi,z=0,t)$, \ quantity \ $\beta(x,y) =
[1+\alpha^2(x,y)]/2$ \ being a bias function chosen as a soft
envelope for the amplitude $\alpha(x,y)$. In order to make easier
the separation of the different diffraction orders from
the encoded complex field $\Psi(\rho,\phi,z,t)$, the off-axis
reference plane wave $\exp[i2\pi(\xi x + \eta y)$ is used, shifting, in the
Fourier plane, the center of signal information to the spatial frequencies $u_0; v_0$.

\h The holographic setup used in the experimental reconstruction
process of the CGH of our ``Lego-beams'' is shown in Fig.4. The
coherent light from the He-Ne laser ($632.8\;$nm) passes through
the spacial filter SF and is collimated by lens L1; it is then
reflected by mirror M1 and polarized by polarizer P1; when
arriving at the beam splitter BS1, the light is directed to the
reflective SLM (model LC-R 1080, Holoeye Photonics), where
the CGH was sent electronically, and diffracted, proceeding to
polarizer P2. To select the correct diffraction order of the
reconstructed (diffracted) beam, a 4F filter is used after the
SLM, composed of two lenses (L2 and L3) and an iris diaphragm
(ID). Finally, the ``Lego-beam'' is acquired by the CCD camera
(model DMK 41BU02.H, Imaging Source), at subsequent locations
along $z$.


\begin{figure}[!h]
\begin{center}
 \scalebox{.7}{\includegraphics{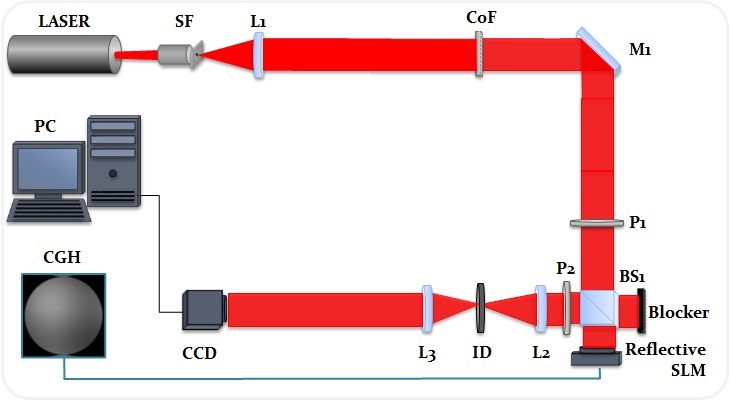}}
\end{center}
\caption{The holographic setup used in the experimental
reconstruction process of the CGH's of the ``Lego-beams''. \
Where: Laser is a He-Ne ($632.8\;$nm) laser; SF is a spatial
filter; L1, L2 and L2 are lens; CoF is a neutral density filter;
M1 is a mirror; P1 and P2 are polarizers; BS1 is a beam splitter;
Blocker is an optical blocker; Reflective SLM is a reflective
Spatial Light Modulator; ID is an iris diaphragm; and, finally,
CCD  is a camera.} \label{fig4}\end{figure}

\h The first experimentally generated beam possesses its shape similar to that of the
first theoretical example, but with different values for the radii of the
cylindrical surfaces due to the limited
resolution of our SLM. Namely, we
generate a beam whose diffraction resistant spatial structure
consists in two adjacent cylindrical surfaces, $12\;$cm and
$16\;$cm long, with the corresponding radii $148\,\mu$m and
$203\,\mu$m, sequentially linked one to the other; while it
coaxially exists a ``light segment" having a spot of $38\,\mu$m
and a length of $28\;$cm.

\h We use, therefore, our ``Lego-beam''-solution, Eq.(\ref{lb}),
with three FWs: one zeroth-order FW, and two eighth-order FWs (the
latter endowed with different values of their central longitudinal
wave numbers). That is, in Eq.(\ref{lb}) the non-zero
functions (FWs) $\psi_{\nu \el}$ are: \ $\psi_{0\,0}$,$\psi_{8\,0}$ \
and \ $\psi_{8\,1}$, \ possessing in the interval $0 \leq z \leq L
= 0.6\;$m the intensity longitudinal patterns:

\bb \begin{array}{clr} F_{\nu l} \,\, =&
1.2\,\delta_{\nu\,0}\,\delta_{\el\,0}\,[H(z-0.2)-H(z-0.48)] \\
\\
& + \delta_{\nu\,8}\,\delta_{\el\,0}\,[H(z-0.2)-H(z-0.32)] \\
\\
& + 1.1\delta_{\nu\,8}\,\delta_{\el\,1}\,[H(z-0.32)-H(z-0.48)]
\,\, ,
\end{array} \label{F1} \ee

where $\delta_{p\,q}$ is a Kronecker delta funcion, and $H(.)$ an
Heaviside function. \ The corresponding $Q_{\nu \el}$ values
depend on the chosen cylidrical surface radii and on the
light-segment spot, and result to be:

\bb \begin{array}{clr} & Q_{0\,0} = 0.9999800\dis{\frac{\om}{c}} \\
\\
& Q_{8\,0} = 0.9999785\dis{\frac{\om}{c}} \\
\\
& Q_{8\,1} = 0.9999885\dis{\frac{\om}{c}}
\end{array} \label{F2} \ee

\h Again, coefficients $A_{\nu \el n}$ are evaluated via
Eq.(\ref{An}); while the longitudinal values, $\beta_{\nu \el n}$,
and the transverse ones,  $h_{\nu \el n}$, of the wave numbers are
given by Eqs(\ref{kz},\ref{h}).  With regard to the number of
terms $N_{\nu \el}$ for each $\psi_{\nu \el}$, we adopted
$N_{0\,0} = 14$, $N_{8\,0} = 10$ and $N_{8\,1} = 10$,
respectively.

\h Figura 5 shows the experimentally generated ``Lego-beam". On
its upper right side, in smaller size, we reproduce its theoretical
prediction.  An excelent agreement is apparent, which confirms
validity and applicability of our method.

\begin{figure}[!h]
\begin{center}
 \scalebox{.12}{\includegraphics{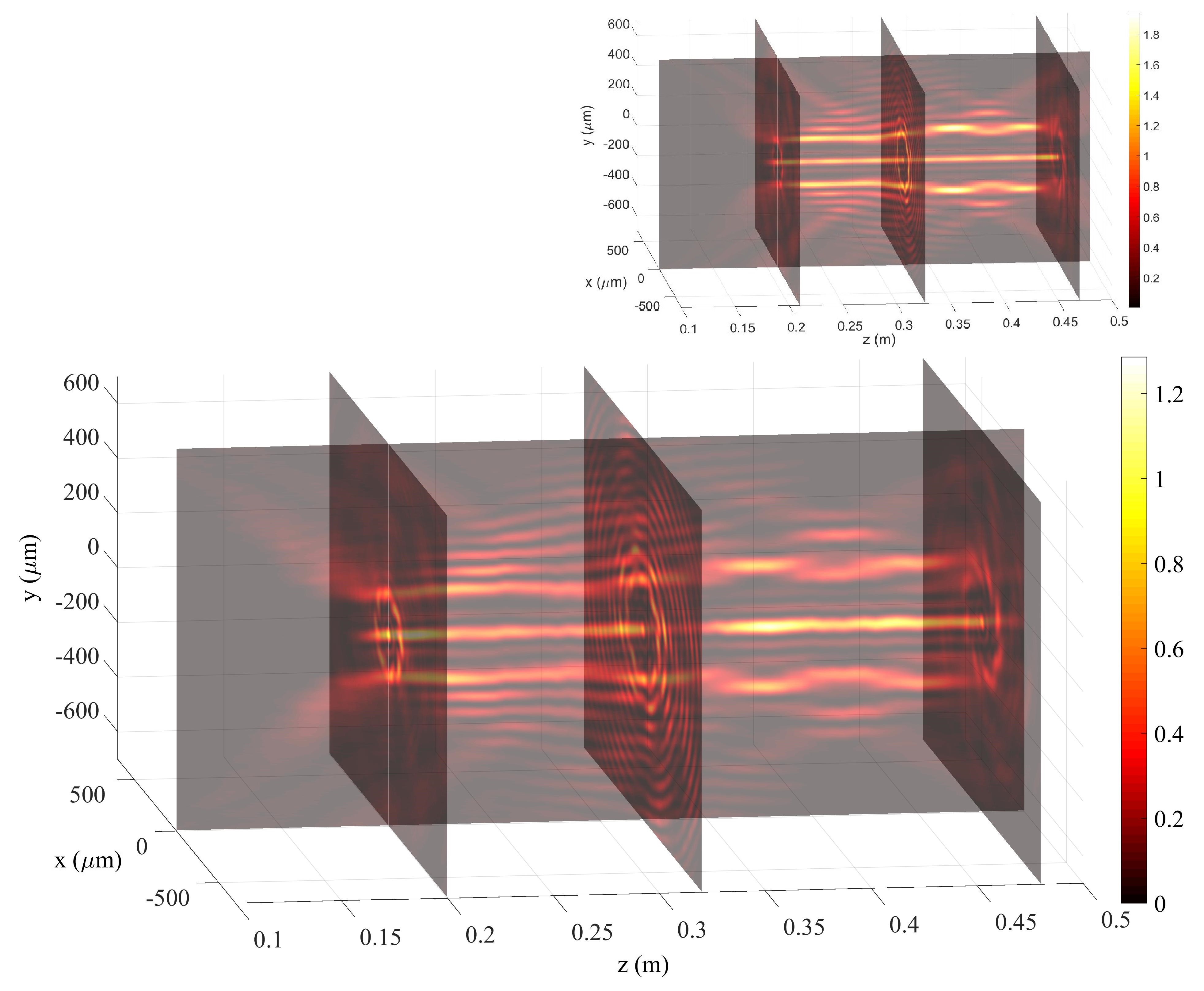}}
\end{center}
\caption{Intensity of the first ``Lego-beam'' generated experimentally,
whose diffraction-resistant spatial structure consists in two
adjacent cylindrical surfaces, $12\;$cm and $16\;$cm long, with
the corresponding radii $148\,\mu$m and $203\,\mu$m, sequentially
linked one to the other; while it exists coaxially also a ``light
segment" having a spot of $38\,\mu$m and a length of $28\;$cm. On
the top right corner, in smaller size, we reproduce the
correspondent theoretical prediction.} \label{fig5}
\end{figure}

\h Let us now pass to the second experimentally generated
``Lego-beam''. We choose two coaxial cylindric light surfaces, in
analogy to our theoretical Fig.3(b), wherein we used two
equal-order FWs (while for the experiment we adopt two
different-order FWs). \ More specifically, we want to generate two
coaxial light-surfaces, having the same length $20\;$cm, but the
different radii $45\,\mu$m and $360\,\mu$m. \ In other words, let
us now adopt the ``Lego-beam''-solution, Eq.(\ref{lb}), with two
FWs only, the first of order 2, and the second of order 22. \
Therefore, in Eq.(\ref{lb}) and in the interval $0 \leq z \leq L =
0.5\;$m, \ the non-zero functions (FWs) $\psi_{\nu \el}$ \ are \
$\psi_{2\,0}$ \ and \ $\psi_{22\,0}$, \ whose longitudinal
intensity patterns are:

\bb \begin{array}{clr} F_{\nu l} \,\, =&
\,\delta_{\nu\,2}\,\delta_{\el\,0}\,[H(z-0.2)-H(z-0.4)] \\
\\
& + \delta_{\nu\,22}\,\delta_{\el\,0}\,[H(z-0.2)-H(z-0.4)]  \,\, .
\end{array} \label{F3} \; , \ee

where, again, $\delta_{p\,q}$ is a Kronecker delta and $H(.)$ a
Heaviside function. \ The corresponding $Q_{\nu \el}$ values,
depending on the chosen cylindrical surface radii, are:

\bb Q_{2\,0} = Q_{22\,0} = 0.9999770\dis{\frac{\om}{c}} \; . \ee

\begin{figure}[!h]
\begin{center}
 \scalebox{.103}{\includegraphics{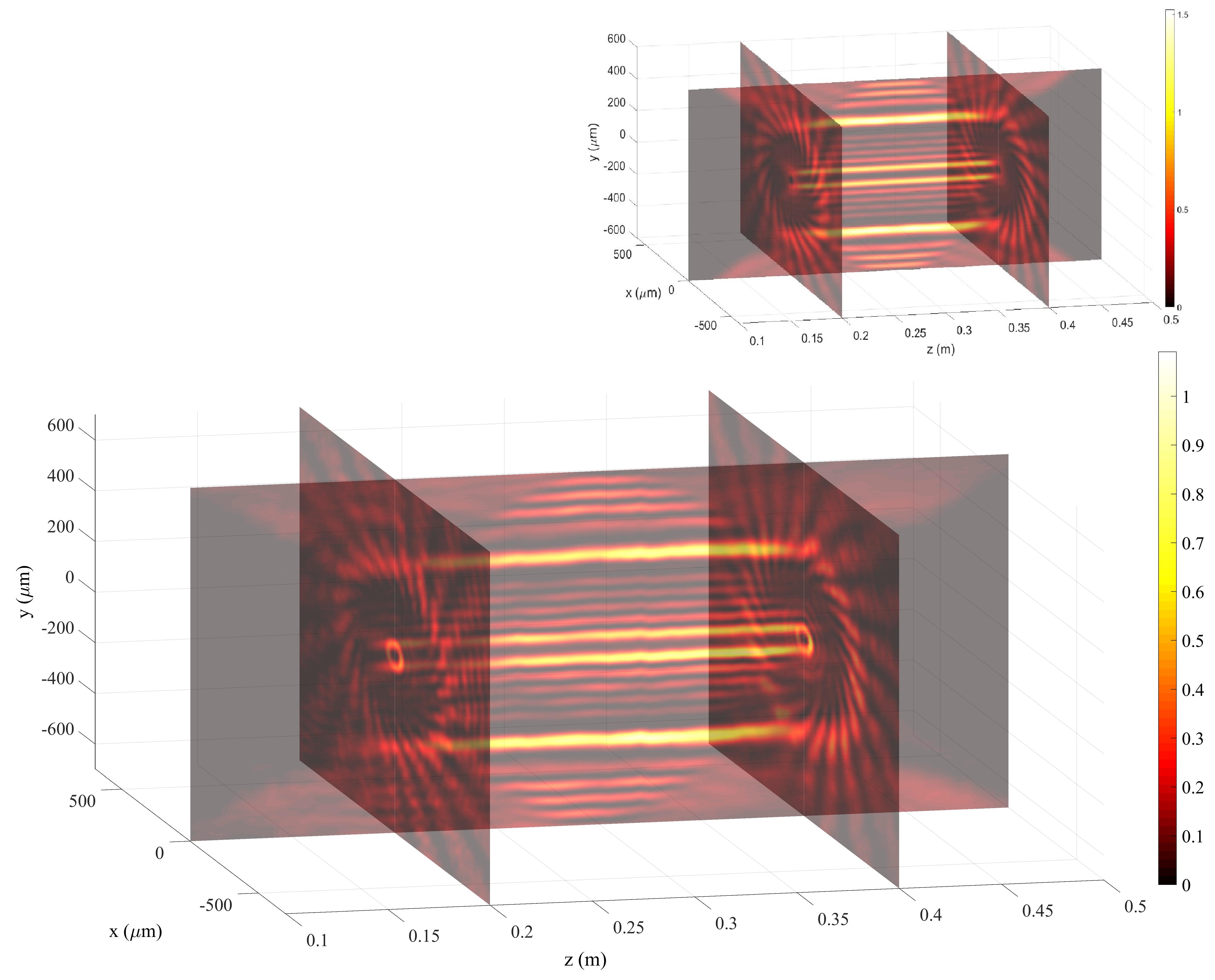}}
\end{center}
\caption{Intensity of the second ``Lego-beam'' generated
experimentally. This beam is composed by two coaxial cylindric
light surfaces having the same length $20\;$cm, but the different
radii $45\,\mu$m and $360\,\mu$m. On the top right corner, in
smaller size, we reproduce the correspondent theoretical
prediction.} \label{fig6}
\end{figure}

As before, coefficients $A_{\nu \el n}$ are evaluated via
Eq.(\ref{An}), and the longitudinal and transverse wave numbers
($\beta_{\nu \el n}$ \ and \ $h_{\nu \el n}$) are calculated by
Eqs(\ref{kz},\ref{h}). For this example we use  $N_{2\, 0} =
N_{22\, 0} = 18$.

\h Figure 6 shows this second experimentally generated beam, and,
on its right side, in smaller size, its theoretical prediction.
Once more, an excellent agreement is found between theory and
experiment; provided that it is kept in mind our discussion of the
interference, between different FWs, present at the end of our
Sec. 2.

In connection with this figure, and with Eq.(\ref{Dp}), let us add
a particular comment: If the external cylinder stays in the region
where $\cos\delta$ is positive (negative), in the resulting beam
its intensity will be higher (lower) than the internal cylinder's.
Such effects can be avoided, or at least reduced, by properly
choosing the (greater or smaller) intensities of the initial
cylinders so that in the final beam they result with the same
intensity.

\section{Conclusions}

\h In this work we present a theoretical method in which ``Frozen
Wave"-type optical beams (of different orders and/or with
different sets of longitudinal wave-number values) are suitably
superposed, providing us with really innovative possibilities
(called by us ``Lego-beams'')  in the important field of
Structured Light.

\h Various theoretical examples are developed and, moreover, our
method has been experimentally verified by the production of two
Lego-beams via a holographic setup, that performs the optical
reconstruction of computer generated holograms through a reflective
Spatial Light Modulator (SLM)
followed by a 4F spatial filtering system.

\h The study of structured light is known to have played an
important role in several areas of optics and photonics; and our
present results can find interesting applications in all sectors
in which more sophisticated light beams are needed, such as
optical tweezers, optical guiding of atoms, light orbital angular
momentum control, imaging systems, remote sensing, light detection
and ranging, microscopy, metrology, optical communications,
quantum information, etc... All these technologies are growing, and
mastering the various types of structured light is becoming more
and more important \cite{dunlop}.

\h For completeness' reasons, let us be here a little more
specific about the expected applications of our ``Lego-beams",
e.g., in the four sectors of optical tweezers for (e.g.)
biological manipulations, atoms guiding, non-linear optics, and
interaction of electromagnetic radiation with Bose-Einstein
condensates:

(i) Optical tweezers are known to be a highly valid instrument for
confining and manipulating nano or micro particles, including
biological ``particles" like bacteria, cells, viruses, etc.  The
use of Non-Diffracting Beams caused many improvements
\cite{arlt,chavez1,macdonald,chavez2,mcgloin,grier,padg,leo1,dolak1,dolak2},
at the point that it was regarded as revolutionary in a Nature's
paper \cite{grier}. Among their advantages, let us mention the
possibility of simultaneously imprisoning many scatterers. Ability
in the spatial modelling of non-diffracting beams, as by our
``Lego-beams", is certainly useful for optical
micro-manipulations, generating new light structures that could
not be imagined before: For instance, by linking together a
zero-order FW (devoid of orbital angular momentum) with a FW of
order 1 or more (carrying orbital angular momentum), it becomes
possible to create a confining region, wherein the particles do
not receive angular momentum, followed by another confining region
in which the particles get angular momentum from the optical field
and start rotating.

(ii) Another sector where use of non-diffracting beams resulted
useful and promising is the optical guiding of neutral atoms, when
Bessel beams of order larger than zero (non-diffracting hollow
beams) generate optical potentials with the same shape
\cite{wang,lemb,arlt2,rhodes,court,dolak3,edwin}. Again, the
possibility of obtaining diffraction-resistant beams with various
interesting, new spatial configurations  (by connecting different
FWs together), namely our  ``Lego-beams", can lead to
unprecedented optical potential configurations for guiding, or
holding, neutral atoms.

(iii) Applications in material modification, more specifically on
the creation of waveguide structures and microchannels for
microfluidics through the process of laser-writing, where a
femtosecond laser induces a material modification (e.g. refractive
index modification) in the longitudinal direction. Studies
\cite{gattas,dharma,zambon,bhuyan} have related good results for
laser micromachining in glass with Bessel beams, obtaining
microchannels with 2 $\mu$m of diameter and with high aspect
ratios (up to 40). ``Lego-beams" could be used for laser-writing
waveguide structures and microchannels (for microfluidics) with a
great variety of forms and, due their diffraction resistant
characteristics, with long longitudinal lengths.

(iv) A fourth interesting sector would be that of examining
(theoretically, and experimentally) the interaction of our
``Lego-beams" with Bose-Einstein condensates.  The
Gross-Pitaevskii equation  --describing a weakly interacting
Bose-Einstein condensate, at the limit of zero temperature-- is
known to possess a mathematical structure similar to the
non-linear Schroedinger equation of optics.  Such a mathematical
equivalence suggests that, while in the non-linear Schroedinger
equation the light is the wave propagating in a medium constituted
of atoms, on the contrary in the Gross-Pitaevsky equation the
atoms play the role of the wave (matter wave) and the light acts
as the propagation medium \cite{cattani}. It is tempting, in this
context, to investigate in theory and experiments the effect of
``Lego-beams" structures on those condensates.

\section{Acknowledgments}

This work is supported by FAPESP (grants 2015/26444-8 and
16/19131-6), CNPQ (grants 304718/2016-5 and  313153/2014-0), and
by CAPES. \ E.Recami thanks CAPES for a visiting professor
fellowship c/o UNICAMP, and Hugo.E.Hern\'andez-Figueroa for his
continuous collaboration and interest.


\begin{thebibliography}{50}

\bibitem{fw1} M.Zamboni-Rached, ``Stationary optical wave fields
with arbitrary longitudinal shape by superposing equal frequency
Bessel beams: Frozen Waves,'' Opt. Express  $\mathbf{12}$(17),
4001--4006 (2004).

\bibitem{fw2} Michel Zamboni-Rached, ``Diffraction-Attenuation resistant beams in
absorbing media," Opt. Express {\bf 14}, 1804-1809 (2006).

\bibitem{fw3} M.Zamboni-Rached, E.Recami, and H.E.Hern\'andez-Figueroa, ``Theory of `frozen waves':
Modeling the shape of stationary wave fields,'' J. Opt. Soc. Am. A
$\mathbf{22}$(11), 2465-2475 (2005).

\bibitem{livro1} {\em Localized Waves}, edited by H.E.Hern\'andez-Figueroa, M.Zamboni-Rached, and E.Recami (J.Wiley; Hoboken, NJ, 2008).

\bibitem{livro2} {\em Non-Diffracting Waves}, edited by
H.E.Hern\'andez-Figueroa, E.Recami, and M.Zamboni-Rached (J.Wiley;
Berlin, 2014).

\bibitem{livro3} {\em Structured Light and Its Applications: An Introduction to
Phase-Structured Beams and Nanoscale Optical Forces}, edited by
David L. Andrews (Academic Press, 2008).

\bibitem{ashkin1} A. Ashkin, J.M. Dziedzic, J.E. Bjorkholm, and S. Chu, ``Observation of a
single-beam gradient force optical trap for dielectric particles",
Opt. Lett.  {\bf 11},  288-290 (1986).

\bibitem{arlt} J. Arlt, V. G\'arces-Ch\'avez, W. Sibbett, K. Dholakia, ``Optical
micromanipulation using a Bessel light beam",  Opt. Commun.
{\bf 197}, 239-245 (2001).

\bibitem{chavez1} V. G\'arces-Ch\'avez, D. McGloin, H. Melville, W. Sibbett, and K.
Dholakia, ``Simultaneous micromanipulation in multiple planes using
a self-reconstructing light beam", Nature {\bf 419}, 145-147 (2002).

\bibitem{macdonald} M. P. MacDonald et al., ``Creation and manipulation of
three-dimensional optically trapped structures", Science {\bf 296}, 1101-1103 (2002).

\bibitem{chavez2} V. G\'arces-Ch\'avez, D. McGloin, M.J. Padgett, W. Dultz, H.
Schmitzer, and K. Dholakia, ``Observation of the transfer of the local
angular momentum density of a multiringed light beam to an
optically trapped particle", Phys. Rev. Lett. {\bf 91}, 093602 (2003).

\bibitem{mcgloin} D. McGloin, V. G\'arces-Ch\'avez, and K. Dholakia, ``Interfering Bessel
beams for optical micromanipulation", Optics Letters {\bf 28}(8), 657-659 (2003).

\bibitem{grier} D.G.Grier, ``A revolution in optical manipulation,'' Nature  $\mathbf{424}$, 810-816 (2003).

\bibitem{padg} M.Padgett, and R.Bowman, ``Tweezers with a twist,'' Nat. Photonics  $\mathbf{5}$(6), 343-348 (2011).

\bibitem{leo1} L.A.Ambrosio, and M.Zamboni-Rached,
``Analytical approach of ordinary frozen waves for optical
trapping and micromanipulation,'' Appl. Opt. $\mathbf{54}$(10),
2584-2593 (2015).

\bibitem{dolak1} G.Milne, K.Dholakia, D.McGloin, K.Volke-Sepulveda, and P.Zem\'anek, ``Transverse particle dynamics in a Bessel beam,'' Opt.
Express {\bf 15}, 13972-13987 (2007).

\bibitem{dolak2} V.Garc\'es-Ch\'avez, D.McGloin, H.Melville, W.Sibbett, and  K.Dholakia, ``Simultaneous micromanipulation in multiple planes
using a self-reconstructing light beam,'' Nature {\bf 419}, 145-147 (2002).

\bibitem{wang} J.Yin, Y.Zhu, W.Wang, Y.Wang, and W.Jhe, ``Optical potential for atom guidance
in a dark hollow laser beam'', J. Opt. Soc. Am. B {\bf 15}, 25-33 (1998).

\bibitem{lembe} V.E. Lembessis, ``A mobile atom in a Laguerre-Gaussian laser beam",
Opt. Commun. {\bf 159}, 243 (1999).

\bibitem{arlt2} J. Arlt, T. Hitomi and K. Dholakia, ``Atom guiding along
Laguerre-Gaussian and Bessel light beams", Applied Physics B {\bf 71}(4),  549-554 (2000).

\bibitem{rhodes} D. P. Rhodes, G.P.T. Lancaster, J. Livesey, D. McGloin, J. Arlt, and
K. Dholakia, ``Guiding a cold atomic beam along a co-propagating
and oblique hollow light guide", Opt. Commun. {\bf 214}, 247-254 (2002).

\bibitem{court} E. Courtade, O. Houde, J.F. Clement, P. Verkerk, and D. Hennequin,
``Dark optical lattice of ring traps for cold atoms", Phys. Rev. A {\bf 74}, 031403 (2006).

\bibitem{dolak3} J.Arlt, T. Hitomi, and K. Dholakia, ``Atom guiding along Laguerre-Gaussian and Bessel light beams,'' Appl. Phys. B
{\em 71}, 549-554 (2000).

\bibitem{edwin} E.G.P.Pachon, M.Zamboni-Rached, A.H.Dorrah,
M.Mojahedi, M.R.R.Gesualdi, and G.G.Cabrera, ``Architecting new
diffraction-resistant light structures and their possible
applications in atom guidance,'' Opt. Express $\mathbf{24}$(22),
25403-25408 (2016).


\bibitem{imaging} T.A.Planchon, L.Gao, D.E.Milkie, M.W.Davidson,
J.A.Galbraith, C.G.Galbraith, and E.Betzig, ``Rapid
three-dimensional isotropic imaging of living cells using Bessel
beam plane illumination,'' Nat. Methods  $\mathbf{8}$(5), 417-423
(2014).

\bibitem{ahmed1} A.H.Dorrah, M.Zamboni-Rached, and M.Mojahedi, ``Controlling the topological charge of twisted
light beams with propagation,'' Phys. Rev. A {\bf 93}, 063864 (2016).

\bibitem{allen1} L. Allen, and M. Padgett, ``Equivalent geometric transformations for
spin and orbital angular momentum of light", J. Mod. Opt. {\bf 54}, 487-491 (2007).

\bibitem{andrews1} D.L. Andrews, L.C.D. Romero, and M. Babiker,`` On optical vortex
interactions with chiral matter", Opt. Commun. {\bf 237}, 133-139 (2004).

\bibitem{araoka} F. Araoka, T. Verbiest, K. Clays, and A. Persoons, ``Interactions of
twisted light with chiral molecules: An experimental
investigation", Phys. Rev. A {\bf 71}, ???? (2005).  

\bibitem{chavez3} V. Chavez, D. McGloin, M.J. Padgett, W. Dultz, H. Schmitzer, and K.
Dholakia, ``Observation of the transfer of the local angular
momentum density of a multiringed light beam to an optically
trapped particle", Phys. Rev. Lett. {\bf 91}, 4 (2003).

\bibitem{livro4}  {\em Paraxial light beams with angular momentum}, by Bekshaev A., and Soskin  
M. (New York: Nova Science Publishers, 2008).


\bibitem{tarcio1} T.A.Vieira, M.R.R.Gesualdi, and M.Zamboni-Rached,
``Frozen waves: experimental generation,'' Opt. Lett.
$\mathbf{37}$(11),  2034-2036 (2012).

\bibitem{tarcio2} T.A.Vieira, M.R.R.Gesualdi, M.Zamboni-Rached, and E.Recami, ``Production of dynamic frozen waves: controlling shape, location (and speed) of diffraction-resistant beams", Opt. Lett.
{\bf 40}(24) 5834-5837 (2015).

\bibitem{mateus} M.Corato Zanarella, and M.Zamboni-Rached,
``Electromagnetic frozen waves with radial, azimuthal, linear,
circular, and elliptical polarizations,'' Phys. Rev. A
$\mathbf{94}$, 053802 (2016).

\bibitem{fw4} M.Zamboni-Rached, and M.Mojahedi,
``Shaping finite-energy diffraction-and attenuation-resistant
beams through Bessel-Gauss beam superposition,'' Phys. Rev. A
$\mathbf{92}$, 043839 (2015).

\bibitem{ahmed2} A.H.Dorrah, M.Zamboni-Rached, and M.Mojahedi,
``Generating attenuation-resistant frozen waves in absorbing
fluid,'' Opt. Lett. $\mathbf{41}$(16), 3702-3705 (2015).

\bibitem{arrizon1} Victor Arriz\'on, ``Optimum on-axis computer-generated hologram
encoded into low-resolution phase-modul ation devices," Opt. Lett.
{\bf 28} 2521-2523 (2003).

\bibitem{arrizon2} Victor Arriz\'on, Guadalupe M\'endez, and David
S\'nchez-de-La-Llave, ``Accurate encoding of arbitrary complex
fields with amplitude-only liquid crystal spatial light
modulators", Opt. Express {\bf 13} 7913-7927 (2005).

\bibitem{dunlop} H.Rubinsztein-Dunlop, {\em et al.}, ``Roadmap on structured light",  J. Opt. {\bf 19}, 013001 (2017).

\bibitem{gattas} R. R. Gattass and E. Mazur, ``Femtosecond Laser Micromachining in
Transparent Materials'', Nat. Photonics {\bf 2}, 219 (2008).

\bibitem{dharma} J. A. Dharmadhikari, A. K. Dharmadhikari, A. Bhatnagar, A.
Mallik, P. Chandrakanta Singh, R. K. Dhaman, K. Chalapathi, and D.
Mathur,  ``Writing low-loss waveguides in borosilicate (BK7) glass
with a low-repetition-rate femtosecond laser", Opt. Commun. {\bf 284},
630 (2011).

\bibitem{zambon} V. Zambon, N. McCarthy, and M. Pich\'e, ``Fabrication of Photonic
Devices Directly Written in Glass Using Ultrafast Bessel Beams",
Proc. SPIE {\bf 7099}, 70992J (2008).

\bibitem{bhuyan} M. K. Bhuyan, F. Courvoisier, P.-A. Lacourt, M. Jacquot, L.
Furfaro, M. J. Withford, and J. M. Dudley, ``High aspect ratio   
taper-free microchannel fabrication using femtosecond Bessel
beams", Opt. Express {\bf 18}, 566 (2010).

\bibitem{cattani} Federica Cattani, Arkady Kim, Mietek Lisak and Dan Anderson,
``Interactions of Electromagnetic Radiation with Bose-Einstein
Condensates: Manipulating Ultra-Cold Atoms with Light",
Int. J. Mod. Phys. B {\bf 27}(6), 1330003 (2013).

\end{thebibliography}
\end{document}